# The superconductivity at 18 K in LiFeAs system

X.C.WANG, Q.Q. LIU, Y.X. LV, W.B. GAO, L.X.YANG, R.C. YU, F.Y.LI, C.Q. JIN*

*Institute of Physics, Chinese Academy of Science, Beijing 100080, China*

**Abstract**

The recent discovery of superconductivity in iron arsenide compounds RFeAsO (R= rare earth) or $AFe_2As_2$ [1~8] (A= alkaline earth) attracts great attention due to the unexpected high $T_c$ in the system containing ferromagnetic elements like Fe. Similar to high $T_c$ cuprates, the superconductivity in iron arsenide is related to a layered structure. Searching for new superconductors with [FeAs] layer but of simpler structure will be of scientific significance either to build up new multilayered superconductors that may reach higher $T_c$ or to study the mysterious underlined superconducting mechanism in iron arsenide compounds. Here we report that a new superconducting iron arsenide system LiFeAs was found. The compound crystallizes into structure containing [FeAs] conducting layer that is interlaced with Li charge reservoir. Superconductivity was observed with $T_c$ up to 18 K in the compounds.

**Corresponding author**: Changqing JIN

Tel: + 86 10 82649163

Fax: + 86 10 82649531

Email: Jin@aphy.iphy.ac.cn

## Introduction

Since the discovery of iron pnictide superconductor[1] with superconducting transition temperature ($T_c$) 26 K for LaFeOAs[2] a series of rare earth iron quaternary pnictide oxides RFeAsO superconductors (R=Ce, Pr. Nd, Sm, …) were found.[3~7] The superconducting transition temperature in ReFeAsO system (or termed "1111" according to the composition ration) was quickly raised above 50K.[6] More recently a new quaternary oxygen free arsenide (Ba, K)$Fe_2As_2$[8] superconductors with $T_c$ 38 K was found that can be alternatively termed as "122" system. Like [CuO2] plane that plays a key role in high $T_c$ superconducting copper oxide[9,10], the [FeAs] layer[11] is believed crucial to support superconductivity in the iron arsenide superconductors. But different from high $T_c$ cuprates that belong to the category of strongly correlated charge transfer type Mott compounds, the layered iron arsenide is an itinerant metal/semi-metal. The unexpected high transition temperature in this itinerant system challenges the conventional BCS mechanism based on electron phonon coupling scenario. Searching for new iron arsenide superconductors with simple structure will be of general interests either to unveil the underlined superconducting mechanism[11-15] or to further enhance $T_c$. Here we report that a new superconducting iron arsenide system LiFeAs (termed "111") was found. LiFeAs crystallizes into a $Cu_2Sb$ type tetragonal structure[16] containing [FeAs] layer with an average iron valence $Fe^{+2}$ like those for "1111" or "122" parent compounds. Superconductivity with $T_c$ up to 18 K was found in the compounds.

## Experimental

The LiFeAs compounds have been synthesized with the assistance of high pressure sintering. The starting materials of Li (99.9%) plus FeAs are mixed according to the nominal formula $Li_{1-z}FeAs$ or $Li_{1+x}FeAs$ ($0 \leq x \leq 0.6$). The FeAs precursors are synthesized from high purity Fe and As powders that are sealed into an evacuated quartz tube. The mixtures are sintered at $800^oC$ for 10 h for several times to assure single phase nature. Since the compositions of LiFeAs are either hygroscopic or easy to react with oxygen or nitrogen, all the process was performed in glove box protected with high purity Ar. We synthesized the "111" type LiFeAs with the assistance of high pressure high temperatures since high pressure can effectively prevent lithium from oxidizing or evaporation upon heating in addition to accelerate the reaction. The pellets of mixed starting materials wrapped with gold foil were sintered at 1GPa to 1.8GPa, $800^oC$ for 60 min followed by quenching from high temperature before releasing pressure. Samples were characterized by x-ray powder diffraction with a Mac Science diffractometer. Diffraction data were collected with 0.02° and 15 s /step. Rietveld analysis has been performed using the program GSAS software package. The magnetic properties of the samples were measured using superconducting quantum interference device (SQUID) (quantum design). The electric conductivity was measured using the standard four probe method.

## Results & Discussions

Figure 1 shows the x-ray diffraction patterns of LiFeAs sample that can be well indexed into a nearly pure $Cu_2Sb$ type tetragonal structure [16] of layered [FeAs] nature with space group P4/nmm. As shown in Fig. 1 the [FeAs] conducting layer is interlaced with Li charge reservoir in LiFeAs. Using Reitveld method the x ray diffraction patterns were also refined based on the structure model [16] as shown in Fig 1 & table 1. This structure is similar to $Fe_2As$ but with the interlayered Fe being replaced with Li. The lattice parameters obtained for LiFeAs are $a$=3.77Å, $c$=6.36Å. Compared with "1111" RFeAsO or "122" $AFe_2As_2$, both $c$ axis or the $ab$ plane is considerably shrunk for "111" LiFeAs. Figure 2 shows the temperature dependence of the electric conductivity of samples with nominal composition $Li_{1-x}FeAs$. The parent LiFeAs shows Pauli metallic behavior with a pronounced curvature at wide temperature range below room temperature, indicating the electron correlation behavior. It could be the intrinsic properties of LiFeAs or it may result from the composition inhomogeniuous that leads to a little bit Li instoichiometry in LiFeAs compounds. It is worth of mentioning that no evidence was observed so far for spin density wave (SDW)-related phase transition in $Li_{1\pm x}FeAs$ system that usually results in resistance drop at the transition temperature as observed for "1111" type RFeAsO or $AFe_2As_2$ (A=Ba, Sr). The electric conductivity was significantly enhanced in samples with nominal composition $Li_{1-x}FeAs$ indicating the possibility for hole doping. The superconducting transition was clearly observed for samples of nominal composition $Li_{1-x}FeAs$ (x=0, 0.2, 0.4) with $T_c$ ranging from 16 K to 18K as shown in Fig.2. The almost constant superconducting transition temperature implies the point superconducting phase feature of Li±FeAs that is in contrast with high $T_c$ cuprates where $T_c$ usually systematically evolves with carrier density. Figure 3 shows the dc magnetic susceptibility of sample with nominal composition Li0.6FeAs measured in both zero field cooling (ZFC) & field cooling (FC) mode with a H=10 Oe. The large Meissner signal (>10%) indicates the bulk superconducting nature of the sample.

Here we address several interesting features for this "111" type LiFeAs superconducting system compared with "1111" type ReFeAsO or "122" type $AFe_2As_2$. There is no evidence so far either from the electric conductivity or magnetic susceptibility measurements showing the SDW for LiFeAs superconducting system that is quite different from "1111" RFeAsO or "122" $AFe_2As_2$ where the superconductivity appears once the SDW transition was suppressed. It is true that the absence of the anomaly cannot completely rule out the presence of SDW order in the parent compound: one possibility is the absence of SDW related structure phase transition observed in "1111" systems [11] that makes the effects on resistivity much smaller. However, if the absence of SDW order in present "111" LiFeAs system is confirmed by further experiments, it will raise an important question: SDW may not be a prerequisite for introducing superconductivity in this [FeAs] compound. Then what is the "glue" leading to superconducting pairing in LiFeAs? The space group of LiFeAs is P4/nmm, i.e. the same as "1111" type RFeAsO so there is no in plane shift for the neighboring interlayer components (charge reservoir layer). On the other hand the alkaline earth layer shifts half periodicity in the *ab* plane for the body-centered "122" type $AFe_2As_2$ with space group I4/mmm. This feature makes LiFeAs more like an "infinite layer structure" comparing with the prototype $ACuO_2$ for high $T_c$ cuprate. [17,18] Consequently the "111"type LiFeAs will be ideal for the theoretical studies due to its plain crystal structure form; or it can be core component to build up multilayered iron arsenide superconductors that may lead to higher $T_c$ as the case for high $T_c$ superconductor cuprates.

Notes: The primary results of this work about the superconducting LiFeAs system have been uploaded as arXiv:0806.4688. Two weeks later other two groups also reported the superconductivity in LiFeAs (arXiv:0807.2228 ; arXiv:0807.2274).

**Acknowledgments:**

The NSF and Ministry of Science & Technology of China are thanked for financial support. We are grateful to Profs. L.Yu, Zhongxian Zhao, & J. S. Zhou for helpful discussions.

**Figure captions**

Fig.1: The powder x ray diffraction pattern as well as Rietveld refinements for a LiFeAs sample crystallizing into a structure that is featured by alternative [FeAs] layer being interlaced with Li similar to the infinite layer $CaCuO_2$.

Fig.2: The temperature dependence of resistivity for LiFeAs compounds showing superconducting transition up to 18K for the sample of nominal composition $Li_{0.6}FeAs$ with the metallic behavior at normal state.

Fig.3: The dc susceptibility of sample with nominal composition $Li_{0.6}FeAs$ in both ZFC & FC mode, indicating the bulk superconducting nature.

**Table 1  Crystallographic data for LiFeAs**

| Atom | site | x | y | z | occupancy |
|------|------|---|---|---|-----------|
| Li | 2c | 0 | 0.5 | 0.343 | 1 |
| Fe | 2a | 0 | 0 | 0 | 1 |
| As | 2c | 0 | 0.5 | 0.735 | 1 |

Then Rietveld refinement of x ray diffraction patterns for LiFeAs at room temperature. Space group: P4/nmm. Unit-cell dimensions: $a=b=3.7715(2)$Å, $c=6.3574(3)$Å; $R_{wp}=8.79\%$, $R_p=5.62\%$, where $R_{wp}$ & $R_p$ are the R factors for the weighted profile, profile, respectively. The refinement range of $2\theta$ was 10-120 °. CuKα radiation was used. The values in parentheses represent the standard deviation in the last digit. The isotropic thermal parameters were set to be constant for all atoms. The occupancy at each site was fixed as 1.

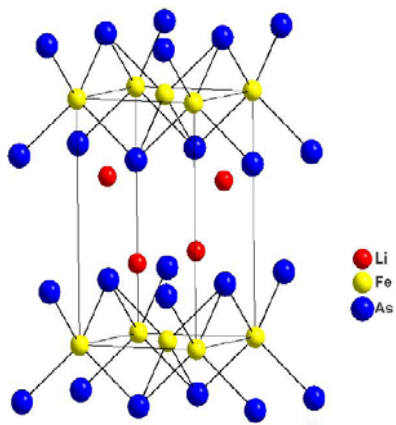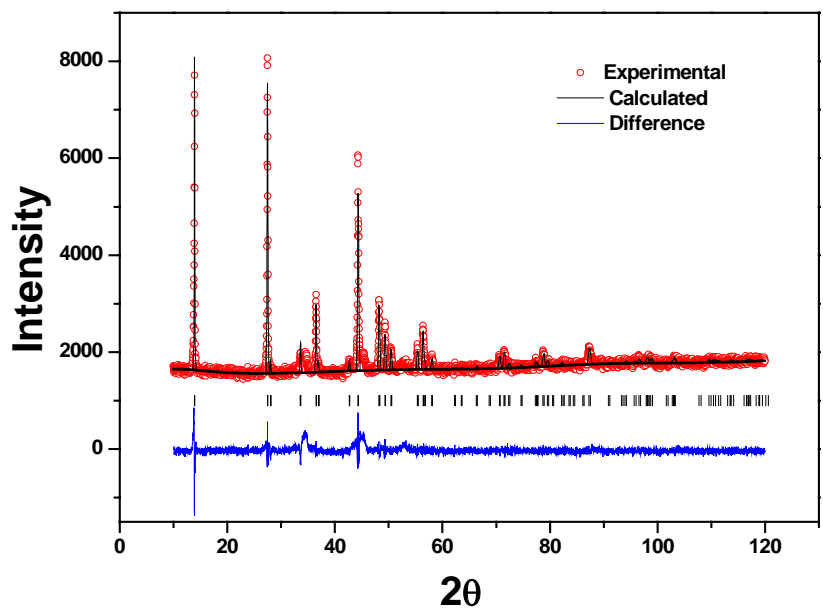

Fig.1

Wang *et al.*

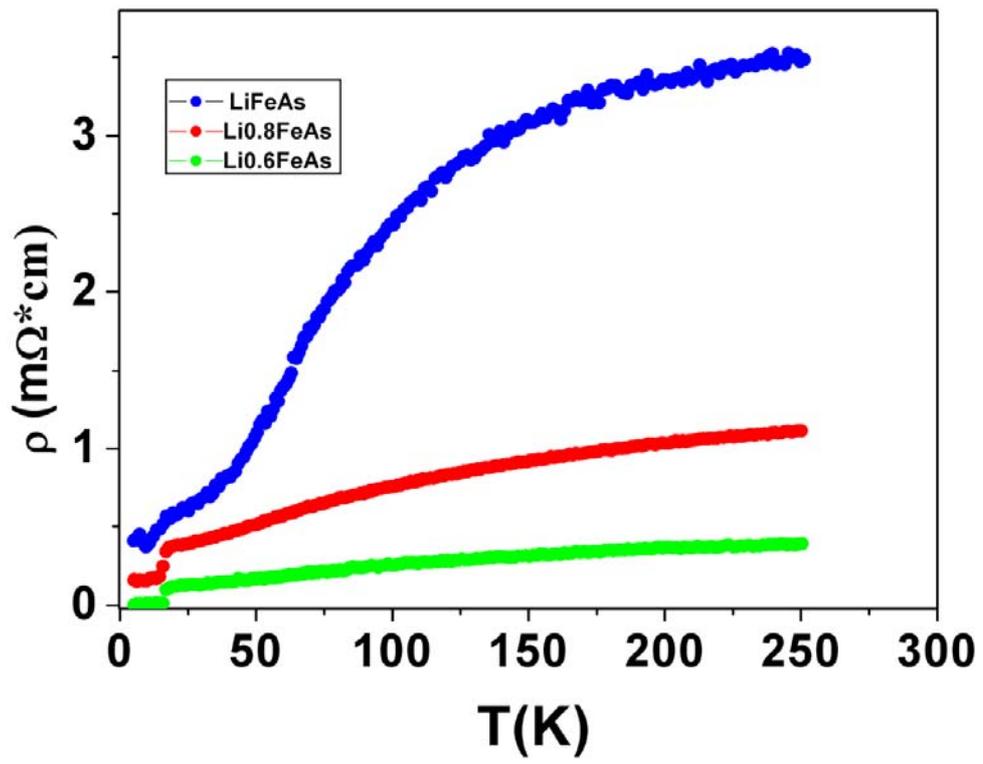

Fig.2

Wang *et al.*

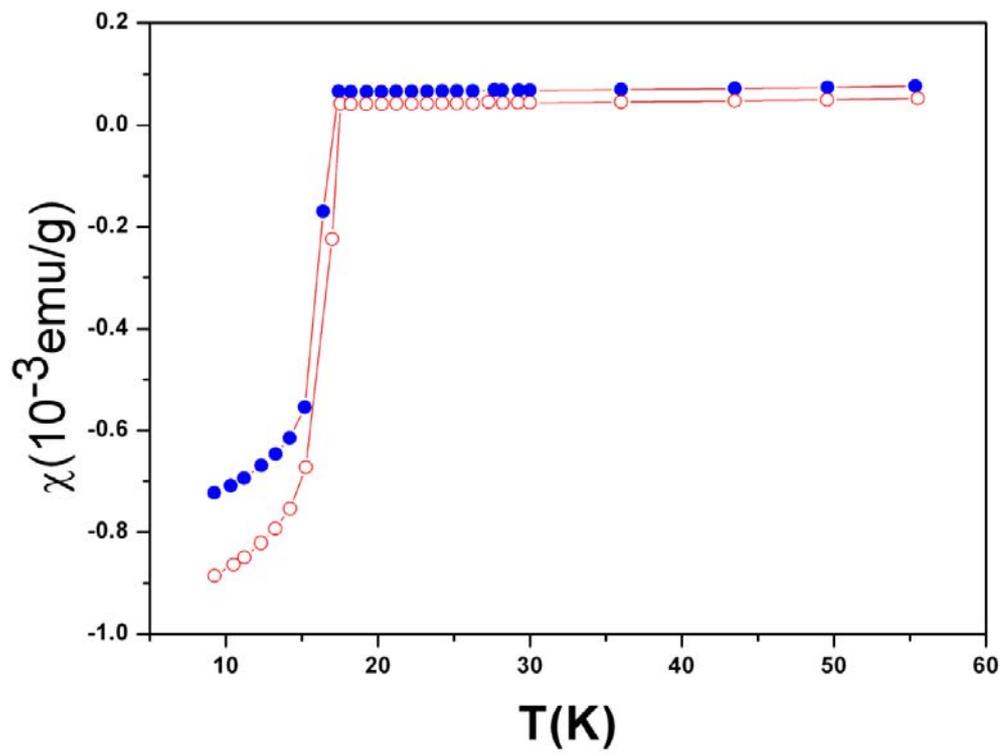

Fig.3

Wang *et al*.